\documentclass[prb,aps,epsf,twocolumn,showpacs,superscriptaddress]{revtex4}
\usepackage[usenames]{color}
\usepackage{graphicx}
\usepackage{epsfig}
\usepackage{latexsym}
\usepackage{amsmath}
\usepackage{color}
\begin{document}

\title{Weak Mott insulators on the triangular lattice: possibility of a gapless nematic quantum spin liquid}

\author{Tarun Grover}
\affiliation{Department of Physics, Massachusetts Institute of
Technology, Cambridge, Massachusetts 02139}
\author{N. Trivedi}
\affiliation{Department of Physics, Ohio State University, Columbus, OH 43210}
\author{T. Senthil}
\affiliation{Department of Physics, Massachusetts Institute of
Technology, Cambridge, Massachusetts 02139}
\author{Patrick A. Lee}
\affiliation{Department of Physics, Massachusetts Institute of
Technology, Cambridge, Massachusetts 02139}

\date{\today}
\begin{abstract}
We study the energetics of Gutzwiller projected BCS states of various symmetries for the triangular lattice antiferromagnet with a four particle ring exchange using variational Monte Carlo methods. In a range of parameters the energetically favored state is found to be a projected $d_{x^2-y^2}$ paired state which breaks lattice rotational symmetry. We show that the properties of this nematic or orientationally ordered paired spin liquid state as a function of temperature and pressure can account for many of the experiments on organic materials. We also study the ring-exchange model with ferromagnetic Heisenberg exchange and find that amongst the studied ans\"atze, a projected $f-$wave state is the most favorable.
\end{abstract}


\newcommand{\be}{\begin{equation}}
\newcommand{\ee}{\end{equation}}
\newcommand{\bea}{\begin{eqnarray}}
\newcommand{\eea}{\end{eqnarray}}
\newcommand{\bK}{\textbf{K}}
\newcommand{\bp}{\textbf{p}}

\maketitle

\section{Introduction}

In the last few years the quasi-two dimensional organic salts
$\kappa-(ET)_2Cu_2(CN)_3$ and $EtMe_3 Sb[Pd(dmit)_2]_2$
(abbreviated respectively as $\kappa CN$ and $DMIT$ in the paper)
have emerged as possible realizations of Mott insulators in the
long sought `quantum spin liquid' state \cite{kanoda03, kanoda05,
kanoda06, kanoda08, kanoda09, kato07, kato08, kato09}. These
layered materials are believed to be well described by the single
band Hubbard model on a nearly isotropic triangular lattice. At ambient
pressure they are Mott insulators which do not order magnetically
down to temperatures $\sim 30 mK$ (much lower than the
exchange $J \sim 250K$ inferred from high temperature
susceptibility) \cite{kanoda03, kato09}. The low temperature phase
is characterized by a linear $T$ dependent heat capacity, and a
finite spin susceptibility just like in a metal (even though the
material is insulating) \cite{kanoda03, kanoda08, kato07}
indicating the presence of low lying spin excitations. There is a
sharp crossover or possibly a phase transition at a low
temperature $\approx 5 K$ signalled by a peak in the heat
capacity, and the onset of a drop in the
susceptibility \cite{kanoda03, kanoda08}. Further an external
magnetic field induces inhomogeneity that is evidenced by a
broadening of the NMR line \cite{kanoda06}. Application of moderate pressure
($\approx 0.5 GPa$) induces a transition to a superconductor ($\kappa CN$) or metal ($DMIT$)
\cite{kanoda05}.

Broadly speaking a spin-liquid ground state of a Mott insulator cannot be smoothly deformed to the ground state of {\em any} electronic band insulator. The theoretical possibility of quantum spin liquids has been appreciated for a long time \cite{anderson73}. Many sharply distinct spin-liquid phases are possible. Further, any quantum spin liquid state possesses exotic excitations with fractional quantum number and various associated topological structures. The distinction between different quantum spin liquid phases is reflected in distinctions of the structure of the low energy effective theory of these excitations.

Currently the most promising candidate materials all seem to share a few key properties. First they are weak Mott insulator that are easily driven metallic by application of pressure. Second they appear to have gapless spin carrying excitations. We are thus lead to study possible gapless spin liquid behavior in weak Mott insulators to understand these materials.

At this point several questions arise: What is a good description of the putative spin liquid Mott state seen in experiments mentioned above?
What is the connection between the superconducting state and the underlying spin liquid state that becomes unstable upon
applying pressure? What is the nature of the finite temperature transitions/crossovers?
We find, using a variational Monte Carlo analysis of the energetics of several possible wave functions for a spin Hamiltonian
with Heisenberg and ring exchange interactions,
that the nodal $d$-wave projected BCS state is the best candidate for the spin liquid. This state has
gapless spin excitations and can naturally explain many of the experiments in $\kappa CN$ though a number of open 
questions remain. We also study the antiferromagnetic $J_4$, ferromagnetic $J_2$ model and find that amongst the studied ans\"atze, a projected $f-$wave state is most favorable. This may have bearing on the explanation of the observed gapless spin-liquid behavior in He-3 films  \cite{masutomi04}.

\subsection{Summary of results}

Our results are based on the model Hamiltonian \cite{motr} 

\bea H & = & 2J_2\sum_{<rr'>} \vec S_r. \vec S_{r'} + J_4 \sum_{\Box}
\left(P_{1234} + h.c\right)  \nonumber \\
& = & J_2 \widetilde{H_2} + J_4 \widetilde{H_4} \label{hmlt}
\eea

 Here $\vec S_r$ are
spin-$1/2$ operators at the sites of a triangular lattice. The
second term sums over all elementary parallelograms, and
$P_{1234}$ performs a cyclic exchange of the four spins at the
sites of the parallelogram. The
multiple ring exchange is expected to be significant due to the
proximity to the Mott transition in the organics. It is known that
the 3 sublattice Neel order vanishes beyond a critical $J_4/J_2\approx 0.1$ \cite{liming} that
can lead to novel spin liquid phases with no long range spin
order.

We study various paired spin liquid states for the $J_2-J_4$
model using variational Monte Carlo calculations. In terms of
wavefunctions, paired states may be described by Gutzwiller
projected BCS states. Two natural states (which retain the full
symmetry of the triangular lattice) are projected singlet $d_{x^2
- y^2} + id_{xy}$ and nodal triplet $f_{x^3-3xy^2}$ wave states. Remarkably in
a range of $J_4/J_2$ with both $J_2$, $J_4$ antiferromagnetic we find that a projected singlet
$d_{x^2-y^2}$ state has better energy than either of these states.
The $d_{x^2- y^2}$ state is a gapless $Z_2$ spin liquid state with
nodal fermionic spinons and gapped $Z_2$ vortices (visons). In
addition it {\em spontaneously} breaks the discrete rotational
symmetry of the triangular lattice but preserves lattice
translational symmetry. Thus it is a gapless $Z_2$ spin liquid
coexisting with a `nematic' or orientational order parameter. 
The pairing structure of the spinons determines the
pairing structure of the superconductor that forms under pressure.
Thus we propose a nodal $d$-wave state for the pressure induced
superconductor as well. Due to the discrete broken rotational symmetry both the insulator and the
superconductor will have non-trivial finite temperature phase
transitions in an ideal sample. We describe these and comment on
their implications for the experiments.

We also study the $J_2-J_4$ Hamiltonian for ferromagnetic $J_2$ while keeping the $J_4$ antiferromagnetic. In this case
we find that of all the states studied, the triplet $f_{x^3-3xy^2}$ wave state has the minimum energy for a large range of $J_4/J_2$. This result may have implications for the gapless spin-liquid behavior observed in 2-D He-3 films \cite{masutomi04}.

\subsection{Relation to earlier Work}

Previous studies on the above model used a Gutzwiller projected filled Fermi sea to interpret
the experiments \cite{motr}. The low energy theory of this state is described
by a gapless Fermi surface of neutral spin-$1/2$
fermionic spinons coupled to a massless $U(1)$ gauge field 
(also obtained \cite{leesq} within a Hubbard model
description). Ref. \cite{motr} concluded that such a state is indeed the minimum energy
state for the $J_2-J_4$ Hamiltonian for $J_4/J_2 \gtrsim 0.30$ but the results were not conclusive for smaller values of
$J_4/J_2$. In particular, Ref. \cite{motr} found that many different projected BCS states in this regime have competitive energies making it difficult to pin-point the true ground state.
One of our goals is to resolve this ambiguity regarding the paired state.
Theoretically, projected BCS states result as a condensation of the spinon pairs that gaps
out the $U(1)$ gauge field, and the resulting state is described
as a $Z_2$ spin liquid. As was pointed out in the Ref.
 \cite{amperean}, such states could explain a sharp crossover observed at $T \sim 5 K$ in the experiments that could be
associated with `pairing' of spinons . Ref. \cite{amperean} suggested an
exotic paired state that retains a finite gapless portion of the
spinon Fermi surface. The possibility of a more conventional
triplet paired $Z_2$ state induced by Kohn-Luttinger effects
\cite{ybkvg} of the spinons has also been pointed out.

Earlier numerical studies on the Hubbard model have also shed  light on the zero-temperature phase
diagram of half-filled triangular lattice \cite{nandini05, watanabe08, tocchio09}. Of course, our model Hamiltonian
(Eqn. \ref{hmlt}) could be thought of as a low-energy limit of a Hubbard model in the insulating regime where we have allowed virtual charge-fluctuations upto four-particle exchange. Using a variational wave-function approach, Ref. \cite{watanabe08} didn't find any evidence for a spin-liquid state and concluded that the insulating regime of the phase diagram is always magnetically ordered. Instead Refs. \cite{nandini05, tocchio09}, using different variational wave-functions, found evidence for a projected nodal $d$-wave wavefunction in the insulating regime in agreement with our result for the ring-exchange Hamiltonian. Though it is not obvious to us which of these studies \cite{nandini05, watanabe08, tocchio09} describe the correct ground state of the half-filled triangular lattice Hubbard model, it is desirable that a theoretical description of the organic salts  $\kappa CN$ or $DMIT$ be consistent with the lack of any apparent order in the insulating regime apart, apart from being able to address other unusual properties mentioned above. We show in this paper that the projected nodal $d$-wave spin liquid state successfully captures many of these features. Since ours is a pure spin model, a direct comparison of the ground state energy with that obtained from variational studies on Hubbard model is not possible.

\subsection{Outline}
The paper is organized as follows. In section \ref{sec:var_wfn} we describe the various variational paired spin liquid states considered in our study along with a brief description of the method of optimization for these variational states. In section \ref{sec:results} we describe the results. In particular, we show that the projected nodal $d_{x^2-y^2}$ state has the minimum energy of all the states considered. In section \ref{sec:expt} we discuss the consequences and predictions of this result in the light of experimental work on organic superconductors. Section \ref{sec:finiteT} considers the finite temperature phase diagram for the nodal $d$-wave superconductor and corresponding projected spin liquid. We conclude with summary and discussion in section \ref{sec:summary}.

\section{Variational wave-functions}
\label{sec:var_wfn}

Various variational states may be constructed by starting with a system
of spin-$1/2$ fermionic spinons $f_{r\alpha}$ hopping on a finite triangular
lattice of size $L1 \times L2$ at half-filling with a ``mean field" Hamiltonian:
\begin{equation}
\label{HMF}
 H_{MF} = \sum_{r r'} \left[  -t_{r r'}f^{\dagger}_{r\sigma} f_{r'\sigma} + \left( \Delta_{rr'} f^{\dagger}_{r\uparrow} f^{\dagger}_{r'\downarrow   } + h.c.\right) \right]
\end{equation}

The variational spin wave-function $|\Psi\rangle_{var} = P_G |\Psi\rangle_{MF}$ where the Gutzwiller projector $P_G = \prod_i \left( 1- n_{i \uparrow} n_{i \downarrow}\right) $ ensures exactly one spinon per site. Unknown parameters in $|\Psi \rangle_{var}$ are fixed by minimizing the energy $E_{var} =  \langle \Psi\,_{var} |H| \Psi \rangle_{var}/  \langle \Psi\,_{var} |\Psi \rangle_{var}$ (with $H$ given by Eq. \ref{hmlt}) with only nearest neighbor $t_{rr'} = t$. The simplest $|\Psi\rangle_{MF}$ corresponds to $\Delta_{rr'} = 0$ i.e. a filled Fermi sea. The corresponding $|\Psi \rangle_{var} \equiv |\Psi \rangle_{PFL}=P_G \left( \prod'_{\vec{k} \sigma} f_{\vec{k} \sigma}^{\dagger} \right) |0 \rangle$ with no variational parameters.
The prime on the product implies restriction to $\vec{k}$ such that the single-spinon level $\epsilon_{\vec{k}} \le E_f$, the Fermi energy.
More complex variational wave-functions are obtained with different patterns of non-zero $\Delta_{rr'}$ which correspond to various projected BCS wavefunctions
$|\Psi \rangle_{PBCS} = P_G |BCS \rangle = P_G \left( \sum_{\vec{k}} \phi_{\vec{k}} f^{\dagger}_{\vec{k} \uparrow} f^{\dagger}_{-\vec{k} \downarrow}\right)^{N/2} |0 \rangle$.
Here $\phi_{\vec k} = \Delta_{\vec{k}}/\left[ \xi_{\vec{k}} + \sqrt{\xi^2_{\vec{k}} + |\Delta_{\vec{k}}|^2}\right] $ with $\xi_{\vec{k}} = \epsilon_{\vec{k}} - \mu$. Further we write $\Delta_{\vec{k}}$, the Fourier transform of $\Delta_{rr'}$, as $\Delta_{\vec{k}} = \Delta_0 F(\vec{k})$
where the form of $F(\vec{k})$ is fully determined from a particular pattern of $\Delta_{rr'}$ (or equivalently, a particular Cooper pair channel).
The two variational parameters: gap parameter $\Delta_0$ and the `chemical potential' $\mu$, are both determined by minimizing the energy. The minimization is done as follows. We first calculate the expectation value of the two-particle exchange term $\widetilde{H_2}$ and the four-particle exchange term $\widetilde{H_4}$ in Eqn. \ref{hmlt} \textit{separately} for a discrete set of values  $(\mu_n,\Delta_{0n})$ such that $\mu_{min} < \mu_n < \mu_{max}$, $\Delta_{min} < \Delta_{0n} < \Delta_{max} $. The ranges for $\mu_{min/max}, \Delta_{0 min/max}$ are chosen based on a preliminary minimization of the energy $E_{var}$ such that the optimum values $\bar{\mu}$, $\bar{\Delta_0}$ lie within this range for \textit{all} value of $J_4/J_2$ we are interested in. Having obtained the discretized functions $\widetilde{H_2}(\mu_{n},\Delta_{0n})$ and $\widetilde{H_4}(\mu_{n},\Delta_{0n})$, the optimization for any  particular value of $J_4/J_2$ is achieved by simply picking the minimum of the function $J_2 \widetilde{H_2}(\mu_{n},\Delta_{0n}) + J_4 \widetilde{H_4}(\mu_{n},\Delta_{0n})$.

\begin{figure}[tb]
\centerline{
 \includegraphics[scale = 0.4]{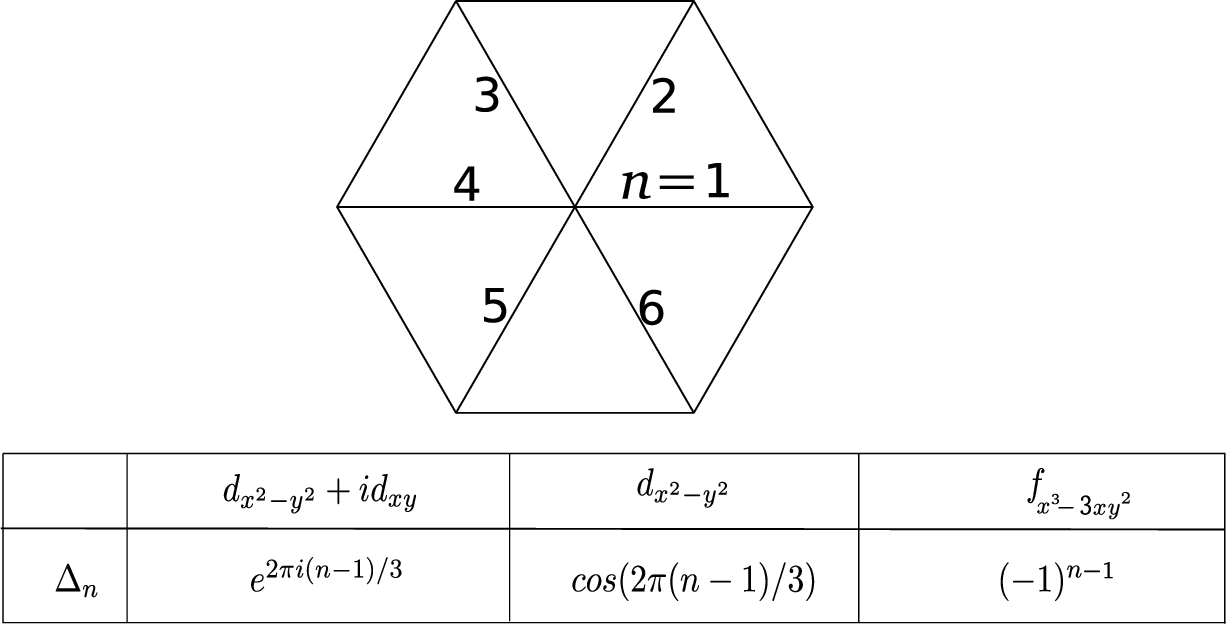}
}
\caption{Angular dependence of superconducting order parameter for the $d_{x^2-y^2}, d_{x^2-y^2} + i d_{xy}$ and $f_{x^3-3xy^2}$ states.}
\label{fig_sc_gap}
\end{figure}

The properties of the three superconducting gap functions are as follows: The $d_{x^2-y^2} + i d_{xy}$ state is invariant under spin rotation, lattice rotation and translation symmetries, but breaks both time-reversal and parity. After projection it corresponds to the `chiral spin-liquid'  state \cite{kalm}. The $d_{x^2-y^2}$ state is a spin-singlet with a $cos(2 \theta)$ angular dependence ($\theta$ is the angle subtended by a bond). It breaks lattice rotational symmetry while preserving translations and time-reversal. Finally, the triplet $f_{x^3-3xy^2}$-wave state has orbital part varying as $cos(3 \theta)$  while in spin-space it has zero projection along a quantization axis. This breaks spin rotation but preserves all the lattice symmetries and time reversal. Both $d_{x^2-y^2}$ and $f_{x^3-3xy^2}$ possess nodes along the Fermi surface in $k$-space while $d_{x^2-y^2} + i d_{xy}$ is fully gapped. Figure \ref{fig_sc_gap} shows the angular dependence of the SC gap $\Delta_{rr'}$ corresponding these three states on the triangular lattice.

\section{Results}
\label{sec:results}

\subsection{Anti-ferromagnetic $J_2$}
\begin{figure}[tb]
\centerline{
 \includegraphics[width=120pt, height=130pt]{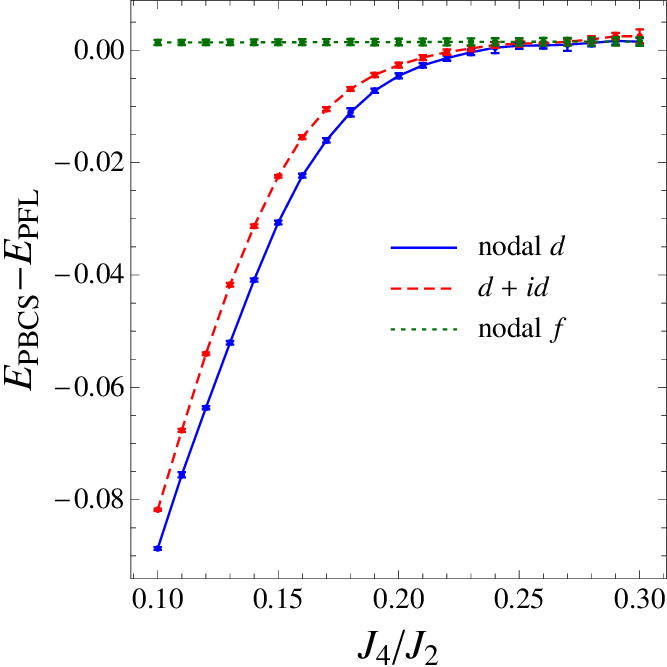}
  \includegraphics[width=120pt, height=130pt]{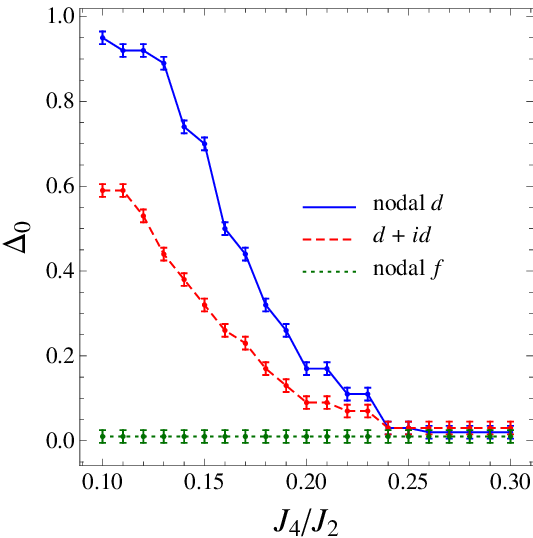}
}
\caption{\underline{Left}: Difference between $E_{PBCS}$ and $E_{PFL}$ for various paired states in the units of $J_2$.
The calculations are done on a $10 \times 11$ lattice with anti-periodic boundary conditions using the standard Metropolis Monte Carlo \cite{gros} with $10^5$ sweeps.
\underline{Right}: Gap parameter $\Delta_0$ for various paired states in units of $t$.}
\label{fig_energy}
\end{figure}

Fig. \ref{fig_energy}(a) shows the the difference $\delta E = E_{PBCS} - E_{PFL}$ for the three paired states, namely projected $d_{x^2-y^2}+id_{xy}$,  $f_{x^3-3xy^2}$ and $d_{x^2-y^2}$. Clearly for $J_4/J_2 \gtrsim 0.25$ the projected Fermi liquid is the best variational state. Interestingly, for a wide range of parameters $ 0.10 \lesssim J_4/J_2 \lesssim 0.23$ the projected $d_{x^2-y^2}$ wins over the projected Fermi liquid as well as the other two paired states. In the regime $ 0.23 \lesssim J_4/J_2 \lesssim 0.25$, the error bars preclude any conclusion. Fig. \ref{fig_energy}(b) shows the optimal value of the gap parameter $\Delta_0$ for these three states. Consistent with the results for optimal energy, $\Delta_0 \approx 0$ for $J_4/J_2 \gtrsim 0.25$ while for $ 0.10 \lesssim J_4/J_2 \lesssim 0.23$ the state $d_{x^2-y^2}$ has a non-zero and largest value of $\Delta_0$ among all paired states. For smaller values of $ J_4/J_2 (\lesssim 0.10$) it is expected that the spin-rotation symmetry breaking spiral state would be the ground state of $H$ \cite{liming}. In addition, we also studied a projected Fermi liquid with staggered flux $\Phi$ through alternate triangular plaquettes. We found that the energy has minima at $\Phi = 0,\pi$ and that the $0$ flux state  $|\Psi \rangle_{PFL}$ is always lower in energy than the $\pi$ flux state for all values of $J_4/J_2$.

Heuristically, large values of $J_4/J_2$ favors delocalization of electrons. Thus it is not surprising that $PFL$ is the ground state for large $J_4/J_2$. Since $J_2 > 0 $, the triplet paired $f$-wave state is expected to be unfavorable, consistent with our results. Further, the electrons are more delocalized in the nodal-$d_{x^2-y^2}$ compared to $d_{x^2-y^2} + id_{xy}$ since the latter is fully gapped. Thus for values of $J_4/J_2$ not so small as to induce spiral order for spins but small enough that $PFL$ is destabilized, our result that a projected nodal paired state is favored  seems reasonable. Our results connect well with earlier
variational Monte Carlo \cite{nandini05} and other numerical studies \cite{kyung} of superconducting states in {\em anisotropic} triangular lattice Hubbard models  
which also found good evidence for a nodal $d$-wave state.

\begin{figure}[tb]
\centerline{
 \includegraphics[width=120pt, height=130pt]{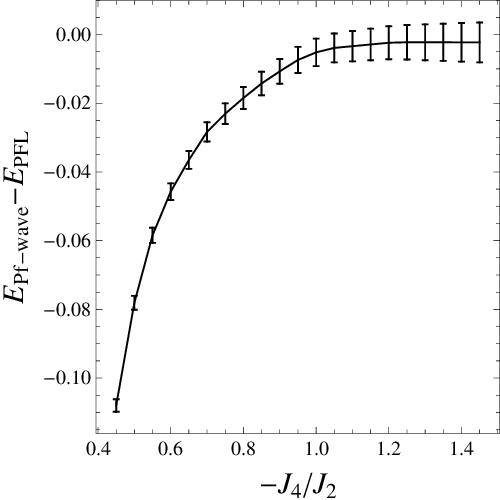}
  \includegraphics[width=120pt, height=130pt]{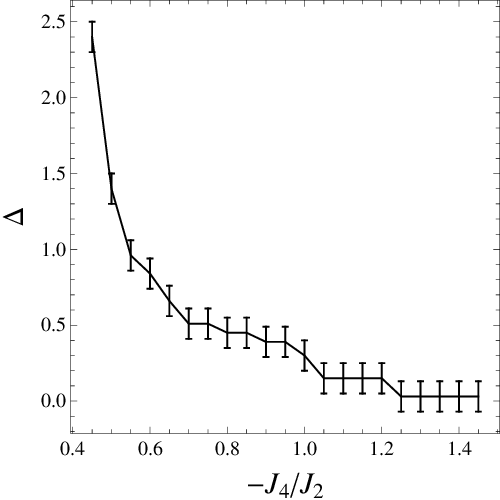}
}
\caption{\underline{Left}: Difference between $E_{Pf-wave}$ and $E_{PFL}$ as a function of $J_4/J_2$ for ferromagnetic $J_2$. Both projected nodal $d$ and $d + id$ have optimal value of $\Delta_0 = 0$ for this sign of $J_2$ and hence are not favorable compared to $PFL$.
\underline{Right}: Gap parameter $\Delta_0$ for projected nodal $f$-wave as a function of $J_4/J_2$.}
\label{fig_ferro}
\end{figure}

\subsection{Ferromagnetic $J_2$}

We also analyze the zero-temperature phase diagram for \textit{ferromagnetic} $J_2$ (Fig. \ref{fig_ferro}). Clearly at $J_4 = 0, J_2 < 0$, one obtains a fully polarized ferromagnet. For any $J_4 > 0$, expectedly we found that the spin-triplet projected nodal $f$-wave state is favored over the spin-singlet projected nodal $d$-wave and $d+id$ states. The projected $f$-wave state becomes favorable also compared to the ferromagnet for $J_4/J_2 \gtrsim 0.40$. Finally for $J_4/J_2 \gtrsim 1.5$, we find that the optimal value of $\Delta \approx 0$ for the projected $f$-wave state and thus the projected FL ($PFL$) state has the lowest energy. This is not very surprising since for $-J_4/J_2 \gg 1$, the sign of $J_2$ shouldn't matter and thus the result is same as that for the anti-ferromagnetic $J_2$ case. Overall, the projected nodal $f$-wave state has the lowest energy of all the states considered (ferromagnet and the projected $d, d+id, f, FL$ states) for $1.5 \gtrsim J_4/J_2 \gtrsim 0.4$.

Under what circumstances is the ferromagnetic $J_2$ ring-exchange model relevant? Technically, a three-particle ring exchange term with strength $J_3 (> 0)$ contributes ferromagnetically to the two-particle exchange because its sole effect is the replacement $J_2 \rightarrow J_2 - 2 J_3$. Thus for $J_3 > J_2/2$, one would obtain a ferromagnetic ring-exchange model. However, a derivation of the ring-exchange Hamiltonian starting from the Hubbard model with only onsite repulsion has $J_n = 0$ for $n$ odd irrespective of the underlying lattice. Since the organic salts  $\kappa CN$ or $DMIT$ are believed to be well-described by the Hubbard model, we believe that the ferromagnetic model is not relevant to their physics. On the other hand, as shown by  Ceperley and Jacucci, \cite{ceper87} $J_3$ is indeed non-zero and bigger than $J_2/2$. Therefore, our results for ferromagnetic $J_2$ ring-exchange model may have implications for the observed spin-liquid behavior in two-dimensional He-3 films \cite{masutomi04}. Further investigation in this direction would be desirable.

\section{Predictions and comparison with experiments}
\label{sec:expt}

We now describe various properties of the state described by the nodal $d$-wave paired wavefunction.
A mean field Hamiltonian that describes the excitations of this state is simply the $H_{MF}$ of Eqn. \ref{HMF}. Fluctuations about the mean field state are described by coupling the spinons to a $Z_2$ gauge field. This state is thus an example of a $Z_2$ spin liquid.
The excitations of the $Z_2$ gauge field
are $Z_2$ flux configurations (known as visons) which are gapped in this spin liquid phase. The low energy physics is then correctly described by the BCS Hamiltonian $H_{MF}$.
Many properties of the nodal $d$-wave spin liquid at low temperature are thus similar to the
familiar {\em spin} physics of a nodal $d$-wave superconductor. We now describe some of these
in relation to the experiments.

\noindent \underline{ {\it Specific heat and spin susceptibility}}:
In the absence of impurities, the density of states for a nodal superfluid vanishes linearly with
energy and consequently the specific heat $C = aT^2$ where the coefficient $a$ is, in principle,
determined by the velocities that characterize the nodal dispersion of the spinons. Impurity scattering
generates a non-zero density of states leading to a specific heat $C \sim \gamma T$, and a constant
spin susceptibility $\chi_0$ as $T \rightarrow 0$. Further the low-$T$ Wilson ratio $\chi T/ C $ is
constant of order one. All of these are in agreement with the experiments on the organic spin liquid materials.

\begin{figure}[tb]
\centerline{
 \includegraphics[scale = 0.25]{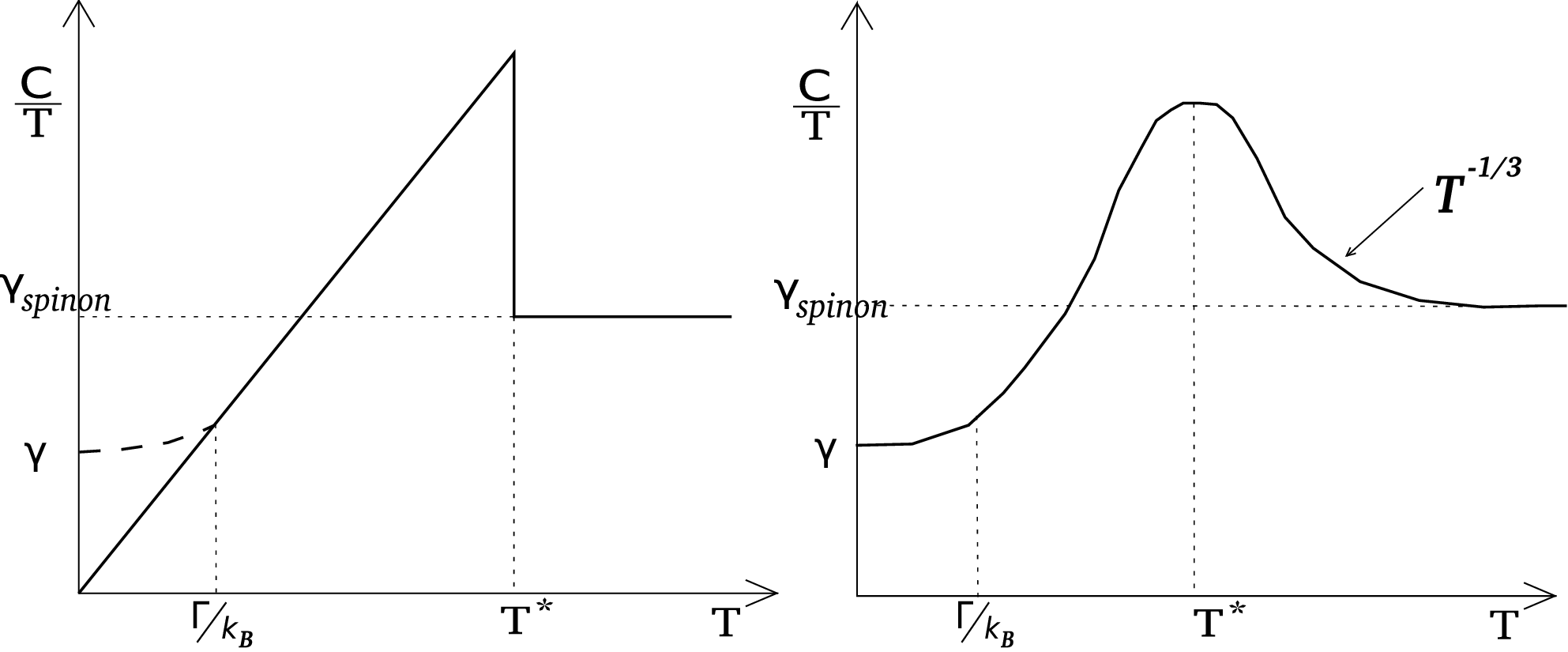}
}
\caption{Schematic sketch of the expected specific heat curve. The left panel shows the ``mean field" behavior. The full line is in the clean limit, and the dashed line the modification due to impurity scattering. The right panel shows the expected behavior when gauge fluctuations beyond mean field are included.}
\label{fig_CT}
\end{figure}

The impurity scattering rate can be roughly estimated by equating
the entropies of the paired nodal spin liquid and `normal' states
at $T^*$, where $T^\ast$ is a mean field or crossover scale below
which the pairing sets in (Fig. \ref{fig_CT}). Above $T^*$ spinons
may be described as having a gapless Fermi surface with a specific heat $C = \gamma_{spinon} T$. Here
$\gamma_{spinon} = (\pi^2/3) k_B^2 n_{spinon}(E_f)$   and
$n_{spinon}(E_f) = 0.28/t_{spinon}$ is the spinon density of
states at the Fermi energy.  Equating the entropies of the paired
and normal states at $T^\ast$, we estimate $a =
2\gamma_{spinon}/{T^*}$. Impurities will cut-off the $T^2$
specific heat of the nodal spin liquid at a scale $\Gamma$ and
lead to a low-$T$ gamma coefficient given by
 $\gamma =a{\Gamma}/{k_B}  = 2\gamma_{spinon} \Gamma/({k_B T^*})$.
Now $t_{spinon}   \lesssim 2J_2 \approx 250$ K and from the
measured low-$T$ specific heat with $\gamma \approx 15 mJ K^{-2}
mol^{-1}$, we estimate an impurity scattering rate $\Gamma
\lesssim 0.25 k_B T^* \approx 1.5K$.  While $\Gamma$ is
reasonably small compared to superconducting gap it is appreciable
enough to generate a constant density of states at low energy and
lead to an apparent Fermi-liquid like behavior in the specific
heat at the lowest temperatures accessible, consistent with the
experiments \cite{kanoda08}. The experiments also apparently show
that the low temperature linear specific heat is insensitive to
magnetic fields upto about $8$ T. This poses a difficulty for the
present theory as the Zeeman coupling to the field is expected to
increase the low energy density of states. Indeed the prior
proposal of an `Amperean' paired state \cite{amperean} was partly motivated by the
insensitivity of the specific heat to a magnetic field. However as
discussed below the Amperean pairing has some difficulty with
describing the superconducting state that develops under pressure.

\noindent \underline{ {\it Thermal conductivity}}:
Nodal spinons  (as also dirty $d$-wave superconductors) lead to a finite `universal' metallic
thermal conductivity $\kappa \sim T$ as $T \rightarrow 0$. In practice however observation of this effect
requires low temperatures to eliminate the phonon contribution. In thermal transport measurements on $\kappa CN$, a plot of $\kappa/T$ as a function of $T^2$ of data above $1$ K
indeed extrapolates to a constant in the zero temperature limit. However, data for $T \lesssim 0.5 $ K rapidly
extrapolates to zero and has been interpreted as evidence for a gap \cite{yamashita09}. We do not have an explanation of this phenomenon.

\noindent \underline{ {\it Field induced inhomogeneity}}:
At ambient pressure NMR studies of $\kappa CN$ show the development of a magnetic-field induced
inhomogeneity \cite{kanoda06}. Within our theory this may be rationalized as follows. Due to the
effect proposed by Motrunich \cite{motr2}, the external magnetic field induces an internal magnetic
field for the spinons which can lead to vortices (visons) of the spinon pair condensate. The
resulting `mixed state' is inhomogeneous that occurs below the pairing scale and increases in proportion to the field.

\noindent \underline{ {\it $T = 0$ phase diagram under pressure
and superconductivity}}: In general pressure increases the  ratio
$t/U$ which implies an increase of $J_4/J_2$ 
leading to suppression of the pair amplitude. Thus
increasing pressure suppresses the pairing transition. Increasing
pressure also leads to an insulator-to-metal transition. Clearly
two situations are possible depending on whether the pair order is
killed before or after this metal-insulator transition. In the
latter case superconductivity will be obtained in the metal close
to the Mott phase boundary. We propose that this is realized in
$\kappa CN$. On the other hand superconductivity has not been
found in $DMIT$ under pressure. We suggest that in this material
the pair order is killed under pressure before the metal-insulator
transition. An interesting experimental test of this suggestion is
to study the Mott insulating phase of $DMIT$ at pressures just
below the metal-insulator phase boundary. Here
 the spinon Fermi surface state, with its characteristic signatures such as, for instance,
 the $T^{2/3}$ heat capacity (produced by gauge fluctuations) will then survive to low-$T$ without any pairing transition.

If the pairing extends into the metallic phase the superconductor
that results will also have $d_{x^2-y^2}$ symmetry, and will (for
an ideal isotropic triangular lattice) break lattice rotational
symmetry. The spinons of the insulator now become the nodal
Bogoliubov quasiparticles of this $d$-wave superconductor. Thus
the low temperature specific heat and spin susceptibilities of the
superconductor will behave similarly to that of the spin liquid
insulator. Further the NMR relaxation rate ${1}/{T_1 T} \sim
T^2$ for $T > \Gamma$ (the impurity scattering rate) and will
saturate to a constant at the lowest temperatures. The former is
in agreement with existing data on $\kappa CN$ for $T$ close to
$T_c$ \cite{kanoda09}.  Such a relaxation rate is not expected within the
alternate Amperean paired state \cite{amperean}, making it difficult to connect
the pairing transition in the spin liquid with that in the metal.
The NMR data \cite{kanoda09} also shows that the Knight shift is only weakly
suppressed on entering the superconducting state. However this may
be due to complications associated with sample heating
\cite{kanoda_private}.

\section{Finite-$T$ phase diagram}
\label{sec:finiteT}

\begin{figure}[tb]
\centerline{
  \includegraphics[width=270pt, height=130pt]{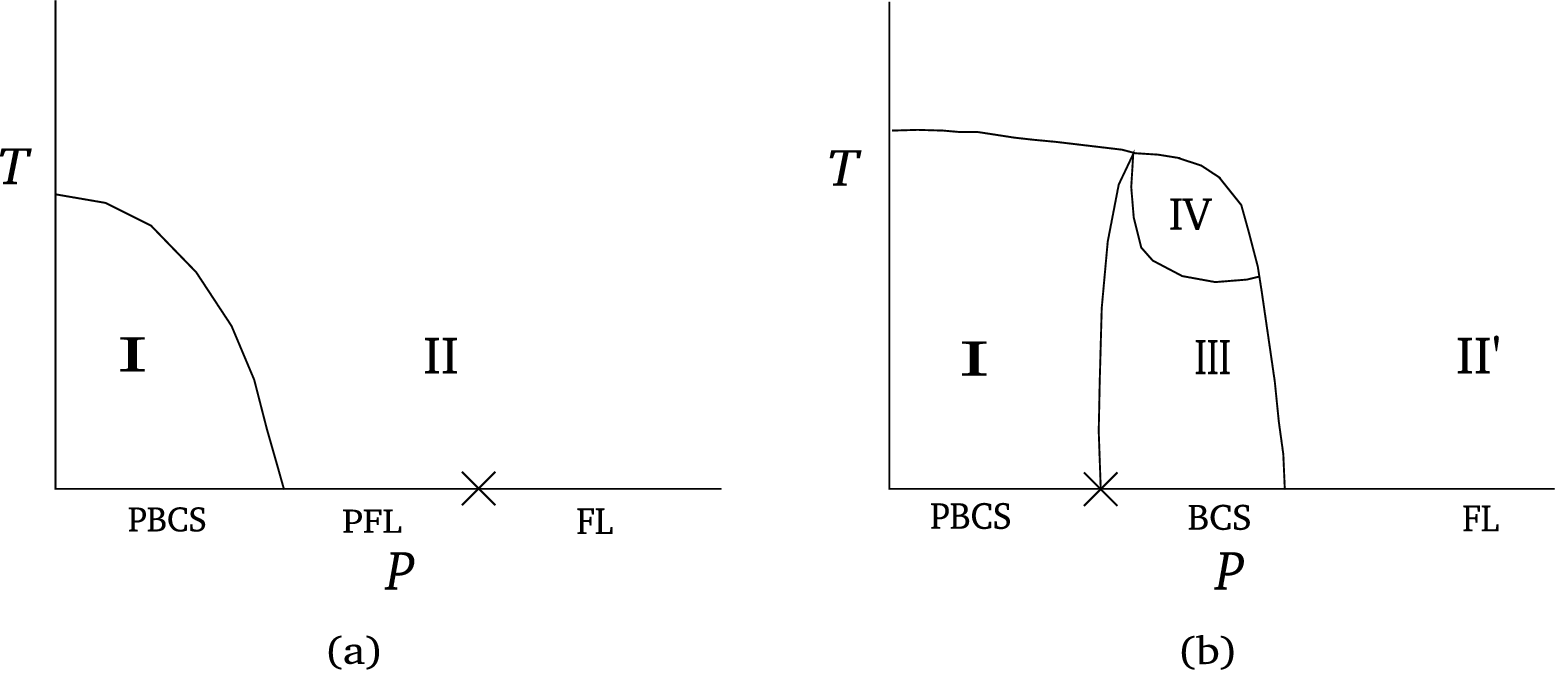}
}
\caption{Schematic pressure-temperature phase diagram for the nodal $d$-wave state. Figure $(a)/(b)$ corresponds to the case when the pair order is killed before/after the Mott transition (denoted by $`\times$'). Phase $I$ corresponds to the finite temperature three-fold symmetry breaking nematic insulator. The details of the regions $II$ and $II'$ are determined by the nature of the Mott transition and/or finite-$T$ crossover phenomena associated with the $FL/PFL$. We do not discuss it in this paper (please see Ref. \cite{senthil08} for a theory of region $II$). The phase $III$ has long-range nematic order and power-law superconducting order while in the region $IV$ both these orders have only power-law correlations. Please see the text for discussion of the associated phase transitions.}
\label{fig_phasedia}
\end{figure}

For an ideal isotropic triangular lattice the broken discrete rotational symmetry of the nodal $d$-wave state leads to an interesting finite temperature phase diagram as shown in Fig. \ref{fig_phasedia}. Let the pairing field $\Delta_{\hat{a}}(\vec r) = \langle c_{\vec{r}\uparrow}c_{\vec {r} + \hat{a}\downarrow} - c_{\vec{r}\downarrow}c_{\vec {r} + \hat{a}\uparrow}\rangle $ on the bond along direction $\hat{a}$ emanating out of the site $\vec r$. Denoting the angle between $\hat{a}$ and the $x$-axis by $\theta_{\hat{a}}$, the order parameter for the nematic superconductor may be written as

\be
\Psi(\vec r) = \sum_{\hat{a}} cos(2\theta_{\hat{a}} - \phi) \Delta_{\hat{a}}(\vec r)
\ee
where the angle $\phi$ describes the orientation of the strong bond of the pair and the sum is over the six possible bond orientations. Since the orientational order is preserved under the symmetry operation corresponding to lattice inversion, $\phi$ takes one of three discrete values $0, 2\pi/3, 4\pi$. Defining
$\psi_{+}(\vec r) = \sum_{\hat{a}} e^{2i\theta_{\hat{a}}}\Delta_{\hat{a}}(\vec r) e^{-i\phi}$, and
$\psi_{-} (\vec r) = \sum_{\hat{a}} e^{-2i\theta_{\hat{a}}}\Delta_{\hat{a}}(\vec r) e^{i\phi}$, we may write
\be
\Psi  = \psi_{+} + \psi_-
\ee

The fields $\psi_{\pm}$ carry electric charge $2$, spin $0$ and transforms non-trivially under lattice and time reversal symmetries. In terms of these new variables, the orientational order parameter is proportional to ${\psi^{*}_{+}} \psi_{-}$. In describing the finite-$T$ phase transition it will be convenient to allow $\phi$ to vary continuously and to impose the discreteness softly through an anisotropy term proportional to $\left({\psi^{*}_{+}} \psi_{-}\right)^3 + c.c$.

First consider the $d_{x^2-y^2}$ paired spin liquid that breaks the discrete orientational symmetry. In the spin-liquid both $\psi_{\pm}$ are coupled to a fictitious U(1) gauge field $\vec{a}$ with gauge charge two. In a phase-only description, one may write $\psi_{+} = e^{i \theta_{1}},\, \psi_{-} = e^{i \theta_{2}}$. A Landau expansion for the free energy $F_I$ consistent with all the symmetries is readily written down in terms of $\theta_{1}, \theta_{2}$ as 

\bea
F_I & =  & \int d^2 \, x \,\, cos(\vec{\nabla} \theta_1 - 2 \vec{a}) + cos(\vec{\nabla} \theta_2 - 2 \vec{a}) \nonumber \\
& & + v \,cos\left( 3 \, (\theta_1-\theta_2)\right) + u \, cos(\nabla \theta_1 - 2 a) cos(\nabla \theta_2 - 2 a)   \nonumber \\
 &  &   + (\vec{\nabla} \times \vec{a})^2
\eea

Choosing the gauge $\theta_2 = 0$, one finds that the above expression for $F_I$ corresponds to an XY model with three-fold anisotropy for the field $\theta_1$. It is known that the critical behavior of this model corresponding to the ordering transition lies in the three state Potts universality class. Thus as the temperature is increased, the insulating nematic phase undergoes a phase transition in the three-state Potts universality class at a certain temperature, say, $T_{c1}$. This corresponds to the phase boundary between phase $I$ and $II'$ in Fig. \ref{fig_phasedia} (b).

The nematic order also leads to a richer finite-$T$ phase diagram in the superconductor. The crucial difference is that now there is no internal gauge field coupled to $\psi_{\pm}$. Thus in the absence of an external electromagnetic field, the Landau free energy $F_{SC}$ may be written as

\bea
F_{SC} & =  & \int d^2 \, x \,\, cos(\nabla \theta_1) + cos(\nabla \theta_2) + v \, cos\left( 3 \, (\theta_1-\theta_2)\right) \nonumber \\
& & + u \, cos(\nabla \theta_1 ) cos(\nabla \theta_2)
\eea

Clearly at temperatures higher than all the energy scales that appear in $F_{SC}$, both $\theta_1$ and $\theta_2$ would be disordered. As the temperature is reduced coherence in $\theta_1, \theta_2$ would start to develop at a certain temperature, say, $T_{c2}$ (which corresponds to the line separating phase $II'$ and $IV$ in Fig. \ref{fig_phasedia} (b)). Let us show that for small $u, v$, both $\theta_1, \theta_2$ undergo Kostrelitz-Thouless (K-T) transitions into a sliding phase at $T_{c2}$. For this to happen, both $u$ and $v$ must be irrelevant at the K-T critical point. As is readily checked, at the K-T fixed point for the decoupled independent variables $\theta_1, \theta_2$, the scaling dimension of $v$ equals $2.25$ while that of $u$ is $4.0$ . Thus both these terms are indeed irrelevant at this fixed point and one concludes that the phase transition is indeed in the 2-d X-Y universality class for both $\theta_1$ and $\theta_2$.

As the temperature is reduced further, the variable $\theta_1 - \theta_2$, which corresponds to the orientational order parameter develops long-range order at some temperature $T_{c3}$, denoted by the line separating the phase $III$ and $IV$ in  \ref{fig_phasedia} (b). To see this, let us introduce the variables $\widetilde{\theta} = (\theta_1 + \theta_2)/2$, $\widetilde{\phi} = (\theta_2 - \theta_1)/2$. In terms of these new variables, the free energy is

\bea
F_{SC} & = & \int d^2 \, x \,\, 2 \, cos(\nabla \widetilde{\phi}) \, cos( \nabla \widetilde{\theta}) +  v\,cos(6 \widetilde{\phi}) \nonumber \\
& & + u/2 \left[ \,cos(2 \nabla \widetilde{\phi}) + cos(2 \nabla \widetilde{\theta})\right] 
\eea

From this one concludes that the action consists of 2-D XY model for $\widetilde{\phi}$  with a 6-fold anisotropy term that is known to be irrelevant at the K-T fixed point. Thus we conclude that the phase transition for the orientational order parameter at the temperature $T_{c3}$ lies in the \textit{inverted} K-T universality class. Note that the phase transition is inverted because the field $\widetilde{\phi}$ has power-law correlations for $ T_{c2}> T > T_{c3}$ while it's connected component $\widetilde{\phi}$ - $\left\langle \widetilde{\phi} \right\rangle$ has exponentially decaying correlations for $T < T_{c3}$.  Thus both transitions at $T_{c2}$ and $T_{c3}$ are in the K-T universality class with very weak signatures in the specific heat. However in both the insulator and the superconductor for the initial pairing transition, by the usual Ginzburg criterion, the fluctuation regime will be rather small. In practice, there is a small lattice anisotropy in both $\kappa CN$ and $DMIT$ which will pin the nematic order parameter and smoothen out any sharp finite temperature nematic transition. Further even in an isotropic material weak disorder acts as a `random field' on the nematic order parameter and will kill the finite temperature nematic phase transition. Nevertheless if the disorder and the lattice anisotropy are weak, a sharp crossover behavior associated with the paired nematic order might be expected in the insulator. Such a crossover is visible in the existing experiments in a variety of properties. In this context we note that recent \textit{ab-initio} calculations \cite{imada09, kandpal09} find an anisotropy $t'/t \sim 0.8$ for $\kappa-CN$. Here the hopping matrix element $t$ corresponds to the parallel bonds of an elementary rhombus on the triangular lattice while $t'$ corresponds to the diagonal bond. Since such an anisotropy would make the lattice symmetry closer to that of a square lattice, physical reasoning as well as the results from papers \cite{nandini05, tocchio09} imply that the nodal $d_{x^2-y^2}$ becomes even more stable for the actual material.

\section{Summary and Discussion}
\label{sec:summary}

To summarize we studied the energetics of various Gutzwiller projected BCS states for the triangular
lattice antiferromagnet with a four particle ring exchange. In a range of parameters the best state
is a projected $d_{x^2-y^2}$ paired state which breaks lattice rotational symmetry. We described many
properties of this state that can account
for a number of measured properties of the organic materials. The
most serious difficulty at present is the field independence of
the specific heat. Perhaps the impurity scattering rate $\Gamma$
can be bigger than our rough estimate without any major effect on
the pairing transition temperature.

Are there any `smoking gun' tests of our proposal for future
experiments? The essence of our proposed gapless $Z_2$ spin liquid
state can be probed in experiments on $\kappa CN$ through the flux
trapping effect described in Ref. \cite{tsmpaf}. We note that such
experiments have been performed on the cuprate materials
\cite{bonn}, and may be feasible in the organics as well. A
Josephson tunneling experiment with a spin liquid insulating
barrier between two superconductors has also been proposed as a
probe of the gapped charge-$e$ spin-$0$ charge carriers expected
in this spin liquid state \cite{tsmpaf2}. Future
experiments will hopefully shed light on whether such a paired nematic spin liquid really exists in these materials. 

We also studied the model with ferromagnetic Heisenberg exchange $J_2$ and anti-ferromagnetic ring exchange $J_4$ and found that amongst the projected BCS/Fermi liquid spin-liquid states, the projected nodal $f$-wave state has minimum energy for wide range of values of $J_4/J_2$.

Finally, we studied the finite temperature phase diagram for the nodal $d$-wave state and found that for an isotropic triangular lattice one would encounter interesting phases and phase transitions as one changes temperature. In particular, there is a possibility of a sliding phase with power-law correlations for both superconducting and orientational order-parameters.

\textit{Acknowledgements}: We thank K. Kanoda, O. Motrunich and Y. Ran for helpful discussions. TS was supported by NSF Grant DMR-0705255, NT by DOE Grant DE-FG02-07ER46423
and PAL by NSF Grant DMR-0804040.

\end{document}